\documentclass[graybox]{svmult}

\usepackage{mathptmx}       
\usepackage{helvet}         
\usepackage{courier}        
\usepackage{type1cm}        
\usepackage{makeidx}         
\usepackage{graphicx}        
\usepackage{multicol}        
\usepackage[bottom]{footmisc}

\makeindex             

\usepackage{amsmath,amssymb,amsfonts}
\usepackage{latexsym,mathrsfs,color}

\def\tr{\mathrm{tr\,}}

\def\im{\mathrm{Im\,}}
\def\ad{\mathrm{ad\,}}

\usepackage{amssymb}
\usepackage{eucal}
\def\tr{\mathrm{tr\,}}

\def\im{\mathrm{Im\,}}
\def\ad{\mathrm{ad\,}}

\def\spn{\mathrm{span\,}}

\def\biglb{\big[\hspace*{-.7mm}\big[}
\def\bigrb{\big]\hspace*{-.7mm}\big]}

\def\openone{\leavevmode\hbox{\small1\kern-3.3pt\normalsize1}}

\def\bbbz{\Bbb{Z}}

\def\m{{\boldsymbol m}}

\def\e{{\bf e}}

\def\ad{\mbox{ad\,}}

\def\tr{\mbox{tr\,}}
\def\im{\mbox{Im\,}}

\def\diag{\mbox{diag\,}}

\def\wedgecomma{\mathop{\wedge}\limits_{'}}

\def\bbbc{{\Bbb C}}
\def\bbbr{{\Bbb R}}

\def\bbbz{{\Bbb Z}}
\def\openone{\leavevmode\hbox{\small1\kern-3.3pt\normalsize1}}

\usepackage{epsf}

\def\newpic#1{%
   \def\emline##1##2##3##4##5##6{%
      \put(##1,##2){\special{em:point #1##3}}%
      \put(##4,##5){\special{em:point #1##6}}%
      \special{em:line #1##3,#1##6}}}

\newpic{}

\def\bbbr{{\Bbb R}}

\def\bbbc{{\Bbb C}}

\def\bbbz{{\Bbb Z}}

\def\wedgecomma{\mathop{\wedge}\limits_{'}}
\def\ad{\mbox{ad}\,}

\def\tr{\mbox{tr}\,}

\def\im{{\rm Im}\,}

\def\fr#1{{\mathfrak{#1}}}
\def\openone{\leavevmode\hbox{\small1\kern-3.3pt\normalsize1}}

\allowdisplaybreaks

\begin{document}

\title*{On nonlocal  models  of Kulish-Sklyanin type and generalized Fourier
transforms}
 \titlerunning{ Nonlocal  models of Kulish-Sklyanin type}

\author{V. S. Gerdjikov}
\institute{V. S. Gerdjikov \at Institute of Nuclear Research and Nuclear Energy,
Bulgarian Academy of Sciences, \\
72 Tsarigradsko chausee, Sofia 1784, Bulgaria, \quad and \\
Institute of Mathematics and Informatics, Bulgarian Academy of Sciences, \\
 Acad. Georgi Bonchev Str., Block 8, 1113 Sofia, Bulgaria \\
\email{gerjikov@inrne.bas.bg}
 }

\maketitle

\abstract{
A special class of multicomponent NLS equations, generalizing the
vector NLS and related to the {\bf BD.I}-type symmetric are shown
to be integrable through the inverse scattering method (ISM). The
corresponding fundamental analytic solutions are constructing thus
reducing the inverse scattering problem to a Riemann-Hilbert
problem. We introduce the minimal sets of scattering data
$\mathfrak{T}$ which determines uniquely the scattering matrix and
the potential $Q$ of the Lax operator. The elements of
$\mathfrak{T}$ can be viewed as the expansion coefficients of $Q$
over the `squared solutions' that are natural generalizations of
the standard exponentials. Thus we demonstrate that the
mapping $\mathfrak{T} \to Q$ is a generalized Fourier transform.
Special attention is paid to two special representatives of this
MNLS with three-component and five components  which describe
spinor ($F=1$ and $F=2$, respectively) Bose-Einstein condensates.}

\section{Introduction}

The integrable multicomponent NLS (MNLS) equations are naturally related to the
symmetric spaces \cite{Man,ForKu*83}. Formally the corresponding MNLS can be written as:
\begin{equation}\label{eq:mnls}\begin{split}
 i \frac{\partial \vec{u}}{ \partial t } + \frac{1}{2} \frac{\partial^2  \vec{u}}{ \partial x^2 } +
\vec{u}  \vec{u}^\dag \vec{u}=0,
\end{split}\end{equation}
see  \cite{TMF98,TMF-144}. Here $\vec{u}$ may be generic $k\times n$ rectangular matrix. It is well known that these
MNLS are related to the A.III class of symmetric spaces in Cartan classification. Of course, one
should consider also the numerous MNLS that can be obtained from (\ref{eq:mnls}) by applying
Mikhailov reductions \cite{Mikh}, see \cite{TMF-144,VSG2,SIGMA-6,GKV,GKV1,1}.

Some of these MNLS have applications to physics. Most of them are related to the vector NLS,
i.e. $k=1$ and  $\vec{u}$ is an $n$-component vector; for $n=2$ this is  the famous Manakov model
\cite{Man}, see also \cite{Konot,dwy08}.

Another very interesting class of MNLS has been discovered by Kulish and Sklyanin  \cite{KuSkl}.
The simplest nontrivial Kulish-Sklyanin (KS) model is a 3-component one
\begin{equation}\label{eq:KS3}\begin{split}
& i\partial_{t} \Phi_{1}+\partial^{2}_{x} \Phi_{1}+2(|\Phi_{1}|^2 +2|\Phi_{0}|^2) \Phi_{1} +2\Phi_{-1}^{*}\Phi_{0}^2=0, \\
& i\partial_{t} \Phi_{0}+\partial^{2}_{x} \Phi_{0}+2(|\Phi_{-1}|^2 +|\Phi_{0}|^2+|\Phi_{1}|^2) \Phi_{0} +2\Phi_{0}^{*}\Phi_{1}\Phi_{-1}=0,\\
& i\partial_{t}\Phi_{-1}+\partial^{2}_{x} \Phi_{-1}+2(|\Phi_{-1}|^2+ 2|\Phi_{0}|^2) \Phi_{-1}+2\Phi_{1}^{*}\Phi_{0}^2=0.
\end{split}\end{equation}
Its integrability,  both in classical and quantum sense, was demonstrated in \cite{KuSkl}, see also \cite{GKV1,SPIE-7501,SPIE-6604}.

The next member in this class is a 5-component one:
\begin{equation}\label{eq:KS5}\begin{split}
&i\partial_t\Phi_{\pm 2}+\partial_{xx}\Phi_{\pm 2}= -2\epsilon ( \Phi, \Phi^{*})(x,t) \Phi_{\pm 2} +\epsilon
\Theta(x,t)  \Phi_{\mp 2}^*, \\
&i\partial_t\Phi_{\pm 1}+\partial_{xx}\Phi_{\pm 1}= -2\epsilon ( \Phi, \Phi^{*})(x,t) \Phi_{\pm 1} -\epsilon
\Theta(x,t) \Phi_{\mp 1}^*,\\
&i\partial_t\Phi_{0}+\partial_{xx}\Phi_{0}=  -2\epsilon ( \Phi, \Phi^{*})(x,t) \Phi_{0} +\epsilon \Theta(x,t)\Phi_{0}^*,
\end{split}\end{equation}
where $\epsilon =\pm 1$ and
\begin{equation}\label{eq:nThe}\begin{split}
( \Phi, \Phi^{*})(x,t) & =\sum_{\alpha=-2}^2 \Phi_{\alpha}\Phi_{\alpha}^{*},\\
\Theta (x,t) &=({{\Phi}},s_0 {{ \Phi}})=2\Phi_{2}\Phi_{-2}-2 \Phi_{1}\Phi_{-1}+\Phi_{0}^2.
\end{split}\end{equation}

Both KS models find important physical applications  in describing spin-1 and spin-2
Bose-Einstein condensates (BEC). Indeed,  BEC of alkali atoms in the $F=1$ hyperfine state,
elongated in $x$ direction and confined in the transverse directions $y,z$ by purely optical means are described by a  3-component
normalized spinor wave vector  $ {\Phi}(x,t)=(\Phi_1, \Phi_0 , \Phi_{-1})^{T}(x,t)$ satisfying the
 equation (\ref{eq:KS3}),  see  \cite{imww04,uiw06,uiw07,dwy08,Kevre*08}:

The assembly of atoms in the  hyperfine state of spin $F$ can be described by a normalized spinor wave vector with $2F+1$
components
\[ {\Phi}(x,t)=(\Phi_{F}(x,t), \Phi_{F-1}(x,t),\dots , \Phi_{-F}(x,t))^{T})^T, \]
whose components are labeled by the values of
$m_F =F, \dots, 1, 0,-1, \dots, -F$. So the spinor BEC with $F=2$ (taken for rather specific choices of the scattering lengths) in dimensionless
coordinates takes the form (\ref{eq:KS5}) \cite{uiw07,GKV1}.
For those who are interested in the physics of spinor BEC we provide some more relevant references
\cite{UK,OM,imww04,uiw06,PRA64,0710a}.

In the last decade  a new trend was started  in nonlinear optics  in attempt to explain artificial heterogenic media.
Such media exhibit  new properties,
 due to the resonance type of interaction of the media and light are observed in photonic crystals, random lasers,
 etc (for a  review, see \cite{UFN}). Some of them can be  modeled by the so-called ${\cal P}$- and
 ${\cal PT}$-symmetric (parity-time) symmetric systems \cite{Konot,Bender1,Bender2, Ali1,Ali2,AbBak,AblMusl,AblMus2,GeSa,Val}.

The initial interest in such systems was motivated by quantum mechanics \cite{Bender1, Ali1}. In \cite{Bender1}
it was shown that quantum systems with a non-hermitian Hamiltonian admit states with real eigenvalues, i.e. the hermiticity
of the Hamiltonian is not a necessary condition to have  real spectrum. Using such Hamiltonians one can build  up new quantum mechanics
\cite{Bender1,Bender2, Ali1,Ali2}.  Starting point is the fact that in the case of a non-Hermitian Hamiltonian with real spectrum,
the modulus of the wave function for the eigenstates is time-independent even in the case of complex potentials.
All this naturally lead  to the development of a special
class of non-local versions of the NLS  equation and its multicomponent versions \cite{TV,AblMusl,GeSa}. The nonlocality
introduced is due to the reductions.

The aim of the present paper is to analyze a special type of  MNLS equations of KS type
and to show that they preserve  integrability also when nonlocal reductions are applied.
For $r=3$ and $r=5$ and with the standard (local) reductions they are characterized by the  Gross-Pitaevsky energy functionals
(see equations (\ref{MFenergy}), (\ref{eq:GPe}) below) and correspond to integrable MNLS models
related to symmetric spaces \cite{ForKu*83} of ${\bf BD.I}$-type
$\simeq {\rm SO(2r+1)}/{\rm SO(2)\times SO(2r-1)}$. Our expose will treat in parallel both reductions.

In Section 2 we give preliminaries about the BEC in one dimension. We also
formulate the Lax representations for  the KS-type equations for any $r$
Section 3 deals with the direct and inverse scattering problem for the Lax operators.
More specifically, we outline the construction of the fundamental analytic solutions (FAS)
of $L$ which allows us to reduce the inverse scattering problem (ISP) for $L$ to a
Riemann-Hilbert problem (RHP). Such approach allows one to use the Zakharov-Shabat dressing
method for calculating the soliton solutions of the KS equations. All these considerations are
valid for Lax operators of generic form, i.e. without any reductions imposed.
 In Section 4 we formulate  the expansions of $q(x,t)$ and its variation $\delta q(x,t)$
over the squared solutions of $L$  for the simplest nontrivial case when $L$ has no discrete eigenvalues.
We will see below, that these expansions are compatible with both the local and non-local $\mathbb{Z}_2$-reductions.
Their expansions coefficients are provided by the minimal sets of scattering data and their variations.
They allow one to generalize  the idea of \cite{AKNS} also to the multicomponent KS-type equations
with both local and nonlocal reductions. Thus we demonstrate that  in all these case the ISM is a
generalized Fourier transform. In  Section 5 we outline the fundamental properties of these NLEE of KS type.
In Section 6 we recall Mikhailov's reduction group which can naturally be applied also to
nonlocal reductions. We derive the constraints on the scattering data imposed by each of these
reductions.

\section{Preliminaries}

\subsection{The BEC in one dimension}

The main tool for investigating BEC is the Gross-Pitaevski (GP) equation
and the GP functional. In the one-dimensional approximation the GP equation in 1D $x$-space  becomes:
\begin{eqnarray}
\label{eq:0} i\frac{\partial{ \Phi}}{\partial t}=\frac{\delta E_{\mathrm{GP}}[{ \Phi}]}{\delta {\Phi^*}}.
\end{eqnarray}
where for $F=1$ the GP energy functional is given by:
\begin{equation}\label{MFenergy}\begin{split}
E_{\rm GP}&=\int d x\, \bigg\{\frac{\hbar^2}{2m}|\partial_x{ \Phi}|^2 +\bar{c}\Big[|\Phi_1|^4+|\Phi_{-1}|^4
+2|\Phi_0|^2(|\Phi_1|^2+|\Phi_{-1}|^2)\Big]  \\
&+(\bar{c}_0-\bar{c}_2)|\Phi_1|^2|\Phi_{-1}|^2 +\frac{\bar{c}_0}{2}|\Phi_0|^4+\bar{c}_2(\Phi^*_1\Phi^*_{-1}\Phi^2_0
+{\Phi^*_0}^2\Phi_1\Phi_{-1})\bigg\}.
\end{split}\end{equation}

For $F=2$ the energy functional  is defined by \cite{OM,UK,uiw07}
\begin{equation}\label{eq:GPe}\begin{split}
E_{\mathrm{GP}}[{ \Phi}]= \int_{-\infty}^\infty d x \left(\frac{\hbar^2}{2m}|\partial_x{ \Phi}|^2
+\frac{\epsilon c_0}{2}n^2 +\frac{c_2}{2}{\bf f}^2+ \frac{\epsilon
c_4}{2}|\Theta (x,t)|^2\right),
\end{split}\end{equation}
where $\epsilon =\pm 1$. The number density $n$ and the singlet-pair
amplitude $\Theta$ are defined in \cite{UK,uiw07}

These two sets of  vector NLS eqs.  can be viewed  as members of  another class of MNLS eqs.
related to the BD.I type of symmetric spaces. They can be written as \cite{SPIE-6604,SPIE-7501,SIGMA-6}:
\begin{equation}\label{eq:KSM}\begin{split}
i \vec{q}_t &+ \vec{q}_{xx} + 2 (\vec{q},\vec{q}^*) \vec{q} -  (\vec{q},s_0\vec{q}) s_0\vec{q}^* =0,
\end{split}\end{equation}
where $\vec{q}$ is $2r-1$-component vector and  the constant matrix $s_0 $ has nonvanishing elements $\pm 1$
only on the second diagonal, see eq. (\ref{eq:z1.6a}) below.

\subsection{Lax Representation for {\bf BD.I}-type MNLS equations }

The symmetric spaces of the series  {\bf BD.I} are isomorphic to $SO(2r+1)/(SO(2)\otimes SO(2r-1) $, see  \cite{Helg}.
The local coordinates on them are provided by the co-adjoint orbits of the algebras $so(2r+1)$ passing through
$J=\diag (1,0,\dots, 0,-1)$. These local coordinates are provided by the matrices
 $q(x,t)=\sum_{\alpha \in \Delta_1^+}( q_\alpha E_\alpha +p_\alpha E_{-\alpha}) $ where the set of roots 
$\Delta_1^+ = \{ e_1 - e_2, \dots , e_1-e_r, e_r, e_1+e_r, \dots ,e_1+e_2 \}$.
For the typical representation we have the  matrix form:
\begin{equation}\label{vec1}
q(x,t)=\left(\begin{array}{ccc}  0 & \vec{q}^{T} & 0 \\
  \vec{p} & 0 & s_{0}\vec{q} \\  0 & \vec{p}^{T}s_{0} & 0 \\
\end{array}\right),\qquad J=\mbox{diag}(1,0,\ldots 0, -1).
\end{equation}
The $2r-1$-component vectors $\vec{q} = (q_2,\dots ,  q_{2r})^T$
are formed by the coefficients  $q_\alpha$  as follows:
$q_k\equiv q_{e_1-e_k}$, $q_{r+1}\equiv q_{e_1}$  and
$q_{2r+1-k}\equiv q_{e_1+e_k}$, $k=2,\dots, r$; the vector
$\vec{p} = (p_2,\dots ,p_{n})^T$ is formed analogously.
The matrix $s_0 =S_0^{(n)}$ enters in the definition of
$so(n)$, i.e. $X\in so(n)$, if $X + S_0^{(n)} X^T S_0^{(n)} =0$,
and for  $n=2r+1$:
\begin{equation}\label{eq:z1.6a}
S_0^{(n)} =  \sum_{s=1}^{n} (-1)^{s+1} E_{s, n+1-s}^{(n)},
\end{equation}
With this definition of orthogonality the Cartan
subalgebra generators are represented by diagonal matrices. By
$E^{(n)}_{sp}$ above we mean $n\times n$ matrix whose matrix
elements are $(E^{(n)}_{sp})_{ij}=\delta_{si}\delta_{pj}$.

The MNLS equations  allow Lax representation $[L,M]=0$ as follows
\begin{eqnarray}\label{eq:3.1}
L\psi (x,t,\lambda ) &\equiv & i\partial_x\psi + (q(x,t) - \lambda J)\psi  (x,t,\lambda )=0.\\ \label{eq:3.2} M\psi (x,t,\lambda )
&\equiv & i\partial_t\psi + (V_0(x,t) + \lambda V_1(x,t) - \lambda ^2 J)\psi  (x,t,\lambda )=0, \\
V_1(x,t)&=& q(x,t), \quad V_0(x,t) = i \ad_J^{-1} \frac{d q}{dx}
+ \frac{1}{2} \left[\ad_J^{-1} q, q(x,t) \right].
\end{eqnarray}

In terms of these notations the generic MNLS type equations connected to ${\bf BD.I.}$ acquire the form
\begin{equation}\label{eq:4.2}
\begin{split}
i \vec{q}_t &+ \vec{q}_{xx} + 2 (\vec{q},\vec{p}) \vec{q} -  (\vec{q},s_0\vec{q}) s_0\vec{p} =0, \\
i \vec{p}_t &- \vec{p}_{xx} - 2 (\vec{q},\vec{p}) \vec{p} + (\vec{p},s_0\vec{p}) s_0\vec{q} =0,
\end{split}
\end{equation}

This equation allows two types of reductions. The first one --
the typical reduction $\vec{p}(x,t)= \vec{q}^{\; *}(x,t)$ is well studied by now, see \cite{KuSkl,ForKu*83,GGK05b}.
The corresponding Hamiltonian for the equations (\ref{eq:4.2}) is given by
\begin{eqnarray}\label{eq:Ham1}
H_{{\rm MNLS}}=\int_{-\infty}^\infty d x \left( (\partial_{x}\vec{q}, \partial_{x}\vec{q^{*}})-
(\vec{q},\vec{q^{*}})^2+ (\vec{q},s_0\vec{q})(\vec{q^{*}},s_{0}\vec{q^{*}})\right),
\end{eqnarray}
For $r=2$  we introduce the variables $\Phi_1=q_{2}$,
$\Phi_{0}=q_3/\sqrt{2}$, $\Phi_{-1}=q_4$; for  $r=3$ we set $\Phi_2 = q_2$,
$\Phi_1 = q_3$, $\Phi_0 = q_4$, $\Phi_{-1} = q_5$ and $\Phi_{-2} =
q_6$. This reproduces the action functionals $E_{GP}$ for $F=1$ and $F=2$.

The second reduction is a non-local one $\vec{p}(x,t)= -\vec{q}^{\; *}(-x,t)$ and is the main topic of
the present paper. As a result we obtain the nonlocal NLS model of {\bf BD.I}-type:
\begin{equation}\label{eq:BD1nls}\begin{split}
i \vec{q}_t &+ \vec{q}_{xx} - 2 (\vec{q}(x,t),\vec{q}^{\; *}(-x,t)) \vec{q}(x,t) +
 (\vec{q}(x,t),s_0\vec{q}(x,t)) s_0\vec{q}^{\; *}(-x,t) =0.
\end{split}\end{equation}

\section{The Direct and the Inverse scattering problem.}\label{sec:2b}

Here we will outline the solution of the direct scattering problem and the construction of the fundamental
analytic solutions (FAS) \cite{Sh}.  The construction goes true for both choices of involutions: local and nonlocal.
Following \cite{ZaSh} we reduce it to a RHP.

\subsection{The Direct scattering problem.}\label{sec:3}

Solving the direct  scattering problem  for $L$ uses the Jost solutions which are defined by, see \cite{VSG2}
and the references therein
\begin{equation}
\lim_{x \to -\infty} \phi(x,t,\lambda) e^{  i \lambda J x }=\openone, \qquad  \lim_{x \to \infty}\psi(x,t,\lambda) e^{i \lambda J x } = \openone
 \end{equation}
and the scattering matrix $T(\lambda,t)\equiv \psi^{-1}\phi(x,t,\lambda)$. The choice of $J$ and the
fact that the Jost solutions and $T(\lambda,t)$ take values in the group $SO(2r+1)$ means that we can use the following block-matrix
structure of $T(\lambda,t)$
\begin{equation}\label{eq:25.1}
T(\lambda,t) = \left( \begin{array}{ccc} m_1^+ & -\vec{B}^-{}^T & c_1^- \\
\vec{b}^+ & {\bf T}_{22} & - s_0\vec{b}^- \\ c_1^+ & \vec{B}^+{}^Ts_0 & m_1^- \\
\end{array}\right), \qquad \hat{T}(\lambda,t) = \left( \begin{array}{ccc} m_1^- & \vec{b}^-{}^T & c_1^- \\
-\vec{B}^+ & s_0{\bf  T}_{22}s_0 &  s_0\vec{B}^- \\ c_1^+ & -\vec{b}^+{}^Ts_0 & m_1^+ \\
\end{array}\right),
\end{equation}
where $\vec{b}^\pm (\lambda,t)$ and $\vec{B}^\pm (\lambda,t)$ are $2r-1$-component vectors, ${\bf T}_{22}(\lambda)$ and
$\m^\pm(\lambda)$ are $2r-1 \times 2r-1$ block matrices, and $m_1^\pm (\lambda)$, $c_1^\pm (\lambda)$ are scalars.
The matrix elements of $T(\lambda,t)$  satisfy a number of  relations which ensure that $T(\lambda)$
belongs to $SO(2r+1)$ and that $T(\lambda)\hat{T}(\lambda) =\openone$. Some of them take the form:
\begin{equation}\label{eq:ThT}\begin{aligned}
m_1^+m_1^- &+ \vec{B}^{-T} \vec{B}^+ + c_1^+c_1^- =1, &\;
\vec{b}^+ \vec{b}^{-T} &+ T_{22}s_0 T_{22}^T s_0 + s_0 \vec{b}^- \vec{b}^{+T} s_0 =\openone, \\
m_1^+m_1^-  &+ \vec{b}^{+T} \vec{b}^- + c_1^+c_1^- =1, &\;
\vec{B}^+ \vec{B}^{-T} &+ s_0T_{22}^Ts_0 T_{22} + s_0 \vec{B}^- \vec{B}^{+T} s_0 =\openone.
\end{aligned}\end{equation}

\subsection{The fundamental analytic solutions}

It is well known that the Jost solutions satisfy a system of Volterra-type integral equations.
Indeed, if we introduce
\begin{equation}\label{eq:JostY}\begin{split}
Y_+(x,t,\lambda) =  \psi(x,t,\lambda) e^{iJ\lambda x}, \qquad  Y_-(x,t,\lambda) =  \phi(x,t,\lambda) e^{iJ\lambda x},
\end{split}\end{equation}
then $Y_\pm (x,t,\lambda)$ must satisfy:
\begin{equation}\label{eq:Y}\begin{split}
Y_{\pm; jk} (x,t,\lambda) = \delta _{jk} + i \int_{\pm \infty}^{x} dy\; e^{-i\lambda (a_j-a_k)(x-y)} \left(
[J,Q(y,t)]Y_\pm (x,t,\lambda)\right)_{jk}.
\end{split}\end{equation}
Here we have used the notation  $J =\diag (a_1,a_2,\dots, a_{2r}, a_{2r+1})$; i.e.  $a_1=1$, $a_2=a_3= \cdots = a_{2r}=0$,
$a_{2r+1}=-1$, (see eq. (\ref{vec1})).

The Volterra equations (\ref{eq:JostY}) always have solution for real $\lambda$. Analytic extension
for $\lambda\in \mathbb{C}_+$ (resp. for $\lambda\in \mathbb{C}_-$) is possible only for the first column
of $Y_- (x,t,\lambda)$ and for the last column of $Y_+ (x,t,\lambda)$ (resp. for the last column
of $Y_- (x,t,\lambda)$ and for the first column of $Y_+ (x,t,\lambda)$. Following Shabat's method \cite{Sh} we
consider two sets of integral equations:
\begin{equation}\label{eq:xi-p}\begin{split}
\xi^+_{jk} (x,t,\lambda) = \delta _{jk} + i \int_{\epsilon_{jk} \infty}^{x} dy\; e^{-i\lambda (a_j-a_k)(x-y)} \left(
[J,Q(y,t)]\xi^+ (y,t,\lambda)\right)_{jk}.
\end{split}\end{equation}
\begin{equation}\label{eq:xi-m}\begin{split}
\xi^-_{jk} (x,t,\lambda) = \delta _{jk} + i \int_{-\eta_{jk} \infty}^{x} dy\; e^{-i\lambda (a_j-a_k)(x-y)} \left(
[J,Q(y,t)]\xi^- (y,t,\lambda)\right)_{jk},
\end{split}\end{equation}
where
\begin{equation}\label{eq:eps}\begin{aligned}
 \epsilon_{jk}=  \begin{cases} 1 &\mbox{for} \quad j\prec k,  \\ -1 &\mbox{for} \quad j \succeq k, \end{cases}, \qquad
 \eta_{jk}=  \begin{cases} -1 &\mbox{for} \quad j \preceq  k,  \\ 1 &\mbox{for} \quad j \succ k, \end{cases}.
\end{aligned}\end{equation}
Here we used the notation
\begin{equation}\label{eq:leq}\begin{aligned}
 j &\prec k &\qquad \mbox{iff} \qquad a_j &>a_k; &\qquad  j &\preceq k &\qquad \mbox{iff} \qquad a_j & \geq a_k; \\
 j &\succ k &\qquad \mbox{iff} \qquad a_j &< a_k; &\qquad  j &\succeq k &\qquad \mbox{iff} \qquad a_j &\leq a_k.
\end{aligned}\end{equation}

Then one can  prove that the equations (\ref{eq:xi-p})  (resp. (\ref{eq:xi-m})) possess solutions $\xi^+ (x,t,\lambda) $
(resp. $\xi^- (x,t,\lambda) $) which allow analytic extension for $\lambda \in \mathbb{C}_+$
(resp. for $\lambda \in \mathbb{C}_-$).
The solutions $\xi^\pm (x,t,\lambda) $ can be viewed also as solutions to a RHP
\begin{equation}\label{eq:xi}\begin{split}
 \xi^+ (x,t,\lambda) = \xi^- (x,t,\lambda) G(x,t,\lambda), \qquad  G(x,t,\lambda) = e^{i\lambda Jx} G_0(t,\lambda)
 e^{-i\lambda Jx}
\end{split}\end{equation}
with canonical normalization, i,e, $\lim_{\lambda \to\infty} \xi^\pm (x,t,\lambda) =\openone$.

If we denote by $ \chi^\pm (x,t,\lambda) = \xi^\pm (x,t,\lambda) e^{-i\lambda Jx}$ then $\chi^\pm (x,t,\lambda) $
will be the FAS of $L$ \cite{Sh,ZMNP,ConMath}. Below we will use two equivalents sets of FAS:
\begin{equation}\label{eq:chi-pm}\begin{aligned}
\chi^\pm (x,t,\lambda) &= \psi(x,t,\lambda) T_J^\mp (t,\lambda) D_J^\pm (\lambda), &\quad
\chi^\pm (x,t,\lambda) &= \phi(x,t,\lambda) S_J^\pm (t,\lambda), \\
\tilde{ \chi}^\pm (x,t,\lambda) &= \phi(x,t,\lambda) S_J^\pm (t,\lambda) \hat{D}_J^\pm(\lambda), &\quad
\tilde{ \chi}^\pm (x,t,\lambda) &= \psi(x,t,\lambda) T_J^\mp (t,\lambda),
\end{aligned}\end{equation}
where $S_J^\pm$, $T_J^\pm$ and $D_J^\pm$ are generalized  Gauss factors of the scattering matrix,
see \cite{ZMNP,G,TMF98,ConMath,1}:
\begin{equation}\label{eq:T}\begin{split}
T(\lambda,t) = T^-_J D^+_J \hat{S}^+_J  = T^+_J D^-_J \hat{S}^-_J ,
\end{split}\end{equation}
where
\begin{equation}\label{eq:25.1'}\begin{aligned}
T^-_J(\lambda,t) &= \left( \begin{array}{ccc}  1 & 0 & 0 \\ \vec{\rho}^+ & \openone & 0 \\ \tilde{c}_1^{-} & \vec{\rho}^{+T}s_0 & 1 \\ \end{array} \right),
&\quad T^+_J (\lambda,t) &=  \left(\begin{array}{ccc}  1 & -\vec{\rho}^{-,T} & \tilde{\tilde{c}}_1^{+} \\ 0 &
\openone & - s_0\vec{\rho}^- \\ 0 & 0 & 1 \\ \end{array} \right), \\
S^+_J(\lambda,t) &= \left(\begin{array}{ccc}  1 & \vec{\tau}^{+T} & \tilde{c}_1^{+} \\ 0 &
\openone & s_0\vec{\tau}^+ \\ 0 & 0 & 1 \\ \end{array} \right), &\quad
S^-_J (\lambda,t)&=  \left( \begin{array}{ccc}  1 & 0 & 0 \\ -\vec{\tau}^- & \openone & 0 \\ \tilde{\tilde{c}}_1^{-} & -\vec{\tau}^{-T}s_0 & 1 \\ \end{array} \right),\\
D^+_J(\lambda) & = \left(\begin{array}{ccc} m_1^+ & 0 & 0 \\ 0 & {\bf m}_2^+ & 0 \\
0 & 0 &  1/m_1^+  \end{array}\right), &\quad
D^-_J(\lambda) & = \left(\begin{array}{ccc} 1/m_1^- & 0 & 0 \\ 0 & {\bf m}_2^- & 0 \\
0 & 0 &  m_1^-  \end{array}\right),
\end{aligned}\end{equation}
We have made use of the following notations above:
\begin{equation}\label{eq:25.1''}\begin{aligned}
\vec{\rho}^\pm &=\frac{\vec{b}^\pm}{m_1^\pm}, &\quad \vec{\tau}^\pm  &=\frac{\vec{B}^\mp}{m_1^\pm}, &\quad
\tilde{c}_1^+ & = \frac{1}{2}( \vec{\tau}^{+T} s_0 \vec{\tau}^+), \\
\tilde{c}_1^- & = \frac{1}{2}( \vec{\rho}^{+T} s_0 \vec{\rho}^+), &\quad \tilde{\tilde{c}}_1^+ & = \frac{1}{2}( \vec{\rho}^{-T} s_0 \vec{\rho}^-),
& \quad \tilde{\tilde{c}}_1^- & = \frac{1}{2}( \vec{\tau}^{-T} s_0 \vec{\tau}^-), \\
c_1^+ &=  \frac{(\vec{b}^{+T} s_0 \vec{b}^+) }{2m_1^+}, &\quad
c_1^- &=  \frac{(\vec{B}^{-T} s_0 \vec{B}^-) }{2m_1^+}.
\end{aligned}\end{equation}

\subsection{The Inverse scattering problem (ISP).}\label{sec:3'}

An important tool for reducing the ISP to a Riemann-Hilbert problem
(RHP) are the fundamental analytic solution (FAS) $\chi^{\pm}(x,t,\lambda )$ and  $\tilde{ \chi}^{\pm}(x,t,\lambda )$.

The Lax representation (\ref{eq:3.1}), (\ref{eq:3.2}) ensures that if $q(x,t) $ evolves according to (\ref{eq:4.2})
then the scattering matrix and its elements satisfy the following linear evolution equations
\begin{equation}\label{eq:evol}
 i\frac{d\vec{\rho}^{\pm}}{d t} \pm \lambda ^2 \vec{\rho}^{+}(t,\lambda ) =0, \qquad i\frac{d\vec{\tau}^{\pm}}{d t}
\mp  \lambda ^2 \vec{\tau}^{\pm}(t,\lambda ) =0, \qquad i\frac{dD ^{\pm}}{d t}  =0,
\end{equation}
so the block-diagonal matrices $D^{\pm}(\lambda)$ can be
considered as generating functionals of the integrals of motion.
The fact that all $(2r-1)^2$ matrix elements of
$\m_2^\pm(\lambda)$ for $\lambda \in \bbbc_\pm$  generate
integrals of motion reflect the superintegrability of the model
and are due to the degeneracy of the dispersion law of
(\ref{eq:4.2}). We remind that $D^\pm_J(\lambda)$ allow analytic
extension for $\lambda\in \bbbc_\pm$ and that their zeroes and
 poles determine the discrete eigenvalues of $L$.

Given the solutions $\chi^\pm(x,t,\lambda)$ one recovers $q(x,t)$ via the formula
\begin{equation}\label{eq:QQ'}
q(x,t) = \lim_{\lambda\to\infty} \lambda \left( J - \chi^\pm J
\widehat{\chi}^\pm(x,t,\lambda)\right).
\end{equation}

The main goal of the dressing method \cite{ZMNP,G,1,I04,GGK05b} is, starting from a known
solutions $\chi^\pm_0(x,t,\lambda)$ of $L_0(\lambda) $ with potential $q_{(0)}(x,t)$ to construct new singular solutions
$\chi^\pm_1(x,t,\lambda )$ of $L$ with a potential $q_{(1)}(x,t)$
with two (or more) additional singularities located at prescribed positions
$\lambda _1^\pm $.  It is related to the regular one by a dressing factor $u(x,t,\lambda )$, for details see \cite{VSG2,gkv08,I04}.

\section{The Generalized Fourier Transforms for non-regular $J$}\label{ch:GFT}

The generalized Fourier transforms (GFT) for the NLEE are based on the completeness relation for the `squared
solutions' of $L$. These completeness relations for the case of generic $J$ have been proved in \cite{G}, see also \cite{TMF98,GeSa}.
In our case  $J$ is highly degenerate: $2r-1$ of its eigenvalues are vanishing. This fact substantially changes the two important steps in
the construction:

i) split the algebra $\mathfrak{g}\simeq so(2r+1)$ into two subspaces: $\mathfrak{g} = \mathcal{O}_J \oplus  \mathcal{O}_J^\perp $. Here
$\mathcal{O}_J $ is the image of the operator $\ad_J$ and provides the co-adjoint orbit in $\mathfrak{g}$ passing through $J$. In our
case $\mathcal{O}_J  \equiv \spn \{ E_\alpha, E_{-\alpha}, \alpha\in \delta_1^+ \}$. $\mathcal{O}_J^\perp $ is the complementary space
orthogonal to $\mathcal{O}_J $ with respect to the Killing form. In what follows we will introduce the operator
$\pi_J = \ad_J^{-1} \ad_J$ which projects any element of $\mathfrak{g}$ onto $\mathcal{O}_J $;

ii) split  each of the `squared solutions' $e_\alpha^\pm (x,\lambda)= \chi^\pm (x,\lambda) E_\alpha \hat{\chi}^\pm(x,\lambda)$ and
$\tilde{e}_\alpha^\pm (x,\lambda)= \tilde{ \chi}^\pm (x,\lambda) E_\alpha \hat{\tilde{ \chi}}^\pm(x,\lambda)$ into two parts:
\begin{equation}\label{eq:e_apm}\begin{aligned}
e_\alpha^\pm (x,\lambda)  = \e_\alpha^\pm (x,\lambda) + e_\alpha^{\pm,\perp} (x,\lambda), \qquad
\tilde{ e}_\alpha^\pm (x,\lambda)  = \tilde{ \e}_\alpha^\pm (x,\lambda) + \tilde{ e}_\alpha^{\pm,\perp} (x,\lambda),
\end{aligned}\end{equation}
where $ \e_\alpha^\pm (x,\lambda)$,  $\tilde{ \e}_\alpha^\pm (x,\lambda)$ belong to $\mathcal{O}_J$,
$ \e_\alpha^{\pm,\perp} (x,\lambda)$ and $\tilde{ \e}_\alpha^{\pm,\perp} (x,\lambda)$ belong to $\mathcal{O}_J^\perp$.

We can view $q(x,t) \in \mathcal{O}_J $ as a generic element of the co-adjoint orbit. The rest of the idea for the GFT is
based on the analyticity properties of the `squared solutions' and on the completeness relation of
$ \e_\alpha^\pm (x,\lambda)$ and $\tilde{ \e}_\alpha^\pm (x,\lambda)$, $\alpha \in \delta_1^+ \cup (-\delta_1^+)$ on
$\mathcal{O}_J$. and is a natural generalization of the proof for generic $J$ \cite{G,TMF98,VSG2}.  Skipping the details we formulate
the expansions for $q(x)$ and $\ad_J^{-1} \delta q(x)$. Of course, for the sake of brevity we treat the case when
the Lax operator $L$ has no discrete eigenvalues.
\begin{equation}\label{eq:49.4}\begin{split}
q(x) &= -{i\over \pi } \int_{-\infty }^{\infty } d \lambda \sum_{\alpha\in\delta_1^+} \left( \tau^+_{\alpha}(\lambda )
\e_{\alpha} ^+(x, \lambda ) -\tau_{\alpha}^-(\lambda ) \e_{-\alpha} ^-(x, \lambda ) \right) \\
&= {i\over \pi } \int_{-\infty }^{\infty } d \lambda \sum_{\alpha\in\delta_1^+} \left( \rho^+_{\alpha}(\lambda )
\tilde{ \e}_{-\alpha} ^+(x, \lambda ) -\rho_{\alpha}^-(\lambda ) \tilde{ \e}_{\alpha} ^-(x, \lambda ) \right) .
\end{split}\end{equation}

\begin{lemma}\label{lem:ms}
Let the potential $q(x,t)$ be such that the Lax operator $L$ has no discrete eigenvalues. Then as minimal set of scattering data
which determines uniquely the scattering matrix $T(\lambda,t)$ and the corresponding potential $q(x,t)$ one can consider either one
of the sets $\mathfrak{T}_i$, $i=1,2$
\[ \mathfrak{T}_1 \equiv \{ \rho_\alpha^+(\lambda,t), \rho_\alpha^-(\lambda,t),  \quad \alpha\in\delta_1^+\}, \qquad
\mathfrak{T}_2 \equiv \{\tau_\alpha^+(\lambda,t), \tau_\alpha^-(\lambda,t),\quad \alpha\in\delta_1^+\}, \]
\end{lemma}
for $\lambda \in \bbbr$.
In other words, the minimal sets of scattering data consist of the expansion coefficients of $q(x)$ over the `squared solutions'.

Similar expansions hold true also for the variation of $q(x)$ \cite{G,TMF98,ConMath}:
\begin{equation}\label{eq:50.6}\begin{split}
\ad_J^{-1}\delta q(x) &= {i \over \pi } \int_{-\infty }^{\infty } d \lambda \sum_{\alpha\in\Delta_1^+} \left(
\delta\tau^+_{\alpha}(\lambda ) \e_{\alpha} ^+(x, \lambda ) +\delta \tau_{\alpha}^-(\lambda )  \e_{-\alpha} ^-(x, \lambda )\right) \\
&= {i \over \pi } \int_{-\infty }^{\infty } d \lambda \sum_{\alpha\in\Delta_1^+} \left(
\delta\rho^+_{\alpha}(\lambda ) \tilde{ \e}_{-\alpha} ^+(x, \lambda ) +\delta \rho_{\alpha}^-(\lambda )  \tilde{ \e}_{-\alpha} ^-(x, \lambda )\right) .
\end{split}\end{equation}
If we consider the special type of variations:
$ \delta q(x) \simeq \frac{\partial q}{ \partial t } \delta t + \mathcal{O}((\delta t)^2) $,
then the expansions (\ref{eq:50.6}) go into
\begin{equation}\label{eq:50.6t}\begin{split}
\ad_J^{-1}\frac{\partial q}{ \partial t} &= {i \over \pi } \int_{-\infty }^{\infty } d \lambda \sum_{\alpha\in\Delta_1^+} \left(
\frac{\partial \tau^+_{\alpha}}{ \partial t } \e_{\alpha} ^+(x, \lambda ) + \frac{\partial \tau_{\alpha}^-}{ \partial t}  \e_{-\alpha} ^-(x, \lambda )\right) \\
&= {i \over \pi } \int_{-\infty }^{\infty } d \lambda \sum_{\alpha\in\Delta_1^+} \left(
\frac{\partial \rho^+_{\alpha}}{ \partial t} \tilde{ \e}_{-\alpha} ^+(x, \lambda ) + \frac{\partial \rho_{\alpha}^-}{ \partial t}  \tilde{ \e}_{-\alpha} ^-(x, \lambda )\right) .
\end{split}\end{equation}

To complete the analogy between the standard Fourier transform and the expansions over the `squared solutions' we need the
generating operators $\Lambda _\pm $:
\begin{equation}\label{eq:**6}
\Lambda _\pm X(x) \equiv \ad_{J}^{-1} \left( i {d X \over d x} + i \left[ q(x), \int_{\pm\infty }^{x} d y\, [q(y), X(y)]\right] \right).
\end{equation}
for which the `squared solutions' are eigenfunctions:
\begin{equation}\label{eq:**0}
\begin{split}
(\Lambda _+-\lambda )\tilde{ \e}_{\mp\alpha}^{\pm} (x,\lambda ) = 0, \qquad
(\Lambda _--\lambda )\e_{\pm \alpha}^{\pm} (x,\lambda ) = 0, \qquad \alpha\in \delta_1^+.
\end{split} \end{equation}

\section{Fundamental properties of the MNLS equations}

The expansions (\ref{eq:49.4}), (\ref{eq:50.6}) and the explicit form
of $\Lambda_\pm$ and eq. (\ref{eq:**0}) are basic for deriving the fundamental
properties of all MNLS type equations related to the Lax operator
$L$. Each of these NLEE is determined by its dispersion law which we choose
to be of the form $F(\lambda) =f(\lambda) J$, where $f(\lambda)$
is polynomial in $\lambda$. The corresponding NLEE becomes:
\begin{equation}\label{nlee}
i\ad_J^{-1} q_t + f(\Lambda_\pm) q(x,t)  = 0.
\end{equation}

\begin{theorem}\label{t1}
The NLEE \eqref{nlee} are equivalent to: i) the equations (\ref{eq:evol}) and ii) the following evolution equations for
the generalized Gauss factors of $T(\lambda)$:
\begin{equation}\label{ds}
i {dS^+_J \over dt} + [F(\lambda), S^+_J] =  0, \qquad i {dT^-_J
\over dt} + [F(\lambda), T^-_J] =  0, \qquad {dD^+_J \over dt}=0.
\end{equation}
or, equivalently. to:
\begin{equation}\label{eq:ropm}
 i\frac{d\vec{\tau}^{\pm}}{d t} \mp f(\lambda)  \vec{\tau}^{\pm}(t,\lambda ) =0, \qquad i\frac{d\vec{\rho}^{\pm}}{d t}
\pm f( \lambda ) \vec{\rho}^{\pm}(t,\lambda ) =0.
\end{equation}
\end{theorem}

The principal series of integrals is generated by the asymptotic
expansion of $\ln m_1^+(\lambda)= \sum_{k=1}^\infty I_k \lambda^{-k}$.
The first integrals of motion are of the form:
\begin{equation}\label{eq:I1-3}
I_1 = -\frac{i}{2} \int_{-\infty }^{\infty } d x\, \langle q(x), q(x) \rangle, \qquad
I_2  = \frac{1}{2} \int_{-\infty }^{\infty } d x\, \langle q_x(x), \ad_J^{-1}q(x) \rangle ,
\end{equation}
Now $iI_1$ can be interpreted as the density of the particles, $I_2$ is the momentum.
The third one  $I_3=iH_{\rm MNLS}$ provides the Hamiltonian.  Indeed,  the
Hamiltonian equations of motion given by $H_{(0)}=-iI_3$ with the Poisson brackets
\begin{equation}\label{eq:PB}
\{ q_{k}(y,t) , p_{j}(x,t)  \} = i \delta _{kj} \delta (x-y),
\end{equation}
coincide with the MNLS equations (\ref{eq:4.2}). The above Poisson
brackets are dual to the canonical symplectic form:
\[ \Omega _0= i \int_{-\infty }^{\infty }d x\, \tr \left(\delta
\vec{p}(x) \wedgecomma \delta \vec{q}(x) \right)=\frac{1}{2i}
\biglb \ad_{J}^{-1} \delta q(x) \wedgecomma \ad_{J}^{-1} \delta q(x) \bigrb, \]
where $\wedgecomma $ means that taking the scalar or matrix product we
exchange the usual product of the matrix elements by wedge-product.

The Hamiltonian formulation of eq. (\ref{eq:4.2}) with $\Omega _0 $ and
$H_0 $ is just one member of the hierarchy of Hamiltonian formulations provided by:
\begin{equation}\label{eq:5.2.6}
\Omega _k = {1 \over i }\biglb \ad_{J}^{-1} \delta Q \wedgecomma
\Lambda ^k \ad_{J}^{-1} \delta Q \bigrb , \qquad  H_k = i^{k+3} I_{k+3}.
\end{equation}
where $\Lambda ={1 \over 2 } (\Lambda_+ +\Lambda _-)$. We can also
calculate $\Omega _k $ in terms of the scattering data variations.
Imposing the reduction $q(x)=q^\dag (x)$ we get:
\begin{equation*}
\begin{split} \Omega _k &=  {1  \over 2\pi i } \int_{-\infty }^{\infty } d\lambda \, \lambda ^k \left( \Omega
_{0}^{+}(\lambda ) -\Omega _{0}^{-}(\lambda ) \right)\\ &=  {1  \over 2\pi}
\int_{-\infty }^{\infty } d\lambda \, \lambda ^k \im \left( m_1^+ (\lambda)
\left( \hat{\m_2}^+ \delta \vec{\rho}^+(\lambda) \wedgecomma \delta \vec{\tau}^+(\lambda)\right) \right).
\end{split}\end{equation*}
This allows one to prove that if we are able to cast $\Omega_{0} $
in canonical form,  then all $\Omega _k $ will also be cast in
canonical form and will be pair-wise equivalent.

\section{The consequences of the involutions}

\subsection{Mikhailov's group of reductions}
The notion of the reduction group  for the integrable NLEE was introduced by Mikhailov in the
beginning of the 1980'ies \cite{Mikh}.

The reduction group $G_R $ is a finite group which preserves the
Lax representation (\ref{eq:3.1}), (\ref{eq:3.2}). This means  that the reduction
constraints are automatically compatible with the evolution. Mikhailov proposed
that  $G_R $ must act on the Lax pair with its two realizations simultaneously:
 i) $G_R \subset {\rm Aut}\fr{g} $ and ii) $G_R \subset {\rm Conf}\, \Bbb C $, i.e. as conformal mappings of the complex
$\lambda $-plane. To each $g_k\in G_R $ we relate a reduction condition for the Lax pair as follows \cite{Mikh}:
\begin{equation}\label{eq:2.1}
C_k(L(\Gamma _k(\lambda ))) = \eta _k L(\lambda ), \quad C_k(M(\Gamma _k(\lambda ))) = \eta _k M(\lambda ),
\end{equation}
where $C_k\in \mbox{Aut}\; \fr{g} $ and $\Gamma _k(\lambda )\in \mbox{Conf\,} \bbbc $
are the images of $g_k $ and $\eta _k =1 $ or $-1 $ depending on the choice of $C_k $. Since $G_R $ is a finite group then for
each $g_k $ there exist an integer $N_k $ such that $g_k^{N_k} =\openone $.
In all the cases below $ N_k=2 $ and the reduction group is isomorphic to $\bbbz_2 $.

More specifically the automorphisms $C_k $, $k=1,\dots,4 $ listed above
lead to the following reductions for the matrix-valued functions
\begin{equation}\label{eq:U-V}
U(x,t,\lambda ) = [J,Q(x,t)] - \lambda J, \qquad V(x,t,\lambda ) = V_0(x,t) +\lambda V_1(x,t) - \lambda^2 J,
\end{equation}
of the Lax representation:
\begin{equation}\label{eq:U-V.a}\begin{aligned}
&\mbox{1)} &\qquad C_1(U^{\dagger}(\kappa _1(\lambda ))) &= U(\lambda ),
&\qquad C_1(V^{\dagger}(\kappa _1(\lambda ))) &= V(\lambda ), \\
& \mbox{2)}  &\qquad C_2(U^{T}(\kappa _2(\lambda ))) &= -U(\lambda ),  &\qquad
C_2(V^{T}(\kappa _2(\lambda ))) &= -V(\lambda ), \\
& \mbox{3)} &\qquad C_3(U^{*}(\kappa _1(\lambda ))) &= -U(\lambda ), &\qquad
C_3(V^{*}(\kappa _1(\lambda ))) &= -V(\lambda ), \\
& \mbox{4)} &\qquad C_4(U(\kappa _2(\lambda ))) &= U(\lambda ), &\qquad C_4(V(\kappa _2(\lambda ))) &= V(\lambda ),
\end{aligned}\end{equation}

 For the nonlocal involutions  we change also $x \to -x$ and find:
\begin{equation}\label{eq:U-V.b}\begin{aligned}
&\mbox{1)} &\qquad C_1(U^{\dagger}(\kappa _1(\lambda ))) &= -U(\lambda ),
&\qquad C_1(V^{\dagger}(\kappa _1(\lambda ))) &= V(\lambda ), \\
& \mbox{2)}  &\qquad C_2(U^{T}(\kappa _2(\lambda ))) &= U(\lambda ),  &\qquad
C_2(V^{T}(\kappa _2(\lambda ))) &= -V(\lambda ), \\
& \mbox{3)} &\qquad C_3(U^{*}(\kappa _1(\lambda ))) &= U(\lambda ), &\qquad
C_3(V^{*}(\kappa _1(\lambda ))) &= -V(\lambda ), \\
& \mbox{4)} &\qquad C_4(U(\kappa _2(\lambda ))) &= -U(\lambda ), &\qquad
C_4(V(\kappa _2(\lambda ))) &= V(\lambda ),
\end{aligned}\end{equation}
Both types of involutions impose constraints on the scattering matrix and on its Gauss factors that are listed below.

\subsection{The local involution case}

The involution:
\begin{equation}\label{eq:locR}\begin{split}
U^\dag (x,t,\kappa_1\lambda^*) &= U(x,t,\lambda) , \qquad \Leftrightarrow \qquad q(x,t) = q^\dag (x,t), \qquad \kappa_1 =1,
\end{split}\end{equation}

On the Jost solutions we have
\begin{equation}\label{eq:locR2}\begin{split}
\phi ^\dag (x,t,\lambda^*) = \phi^{-1}  (x,t,\lambda) , \qquad  \psi ^\dag (x,t,\lambda^*) = \psi^{-1}  (x,t,\lambda) ,
\end{split}\end{equation}
so for the scattering matrix we have
\begin{equation}\label{eq:Tdag}\begin{split}
 T ^\dag (t,\lambda^*) = T^{-1}  (t,\lambda) ,
\end{split}\end{equation}
and for the Gauss factors:
\begin{equation}\label{eq:Spm}\begin{aligned}
 S^-{}^\dag (\lambda^*) & = \hat{S}^+(\lambda), &\quad  T^-{}^\dag (\lambda^*) & = \hat{T}^-(\lambda), &\quad
 D^-{}^\dag (\lambda^*) & = \hat{D}^+(\lambda),
\end{aligned}\end{equation}
Note that the FAS can be used to define the kernel of the resolvent of $L$ by
$R^\pm (x.y.\lambda) =-i \chi^\pm(x,\lambda) \Theta^\pm (x-y) \hat{\chi}^\pm(y,\lambda)$, where the functions
$\Theta^\pm (x-y)$ satisfy the equation $\frac{\partial }{ \partial x }\Theta^\pm (x-y)=\delta(x-y) \openone$ \cite{ConMath,VSG2}.
Next, one can fix up $\Theta^\pm (x-y)$ in such a way that $R^\pm(x,y,\lambda)$ fall off exponentially for $x,y\to \pm \infty$.
So, if $D^+(\lambda)$ (or $D^-(\lambda)$) have a zero or a pole at $\lambda=\lambda_1^+$ (or  at $\lambda=\lambda_1^-$)
then $\lambda_1^\pm$ will be poles of $R^\pm(x,y,\lambda)$ and consequently, discrete eigenvalues of $L$.

If we have local reduction, then
\begin{equation}\label{eq:tau}\begin{split}
\tau^+(\lambda) = -\tau^{-,*} (\lambda), \qquad \rho^+(\lambda) = -\rho^{-,*} (\lambda),
\end{split}\end{equation}

\subsection{The nonlocal involution case}

Now the involution is:
\begin{equation}\label{eq:nlocR}\begin{split}
U^\dag (x,t, \lambda^*) &= -U(-x,t,-\lambda) , \qquad \Leftrightarrow \qquad  q(x,t) = q^\dag (-x,t).
\end{split}\end{equation}
On the Jost solutions we have
\begin{equation}\label{eq:nlocR2}\begin{split}
\phi ^\dag (x,t,\lambda^*) = \psi^{-1}  (-x,t,-\lambda) , \qquad  \psi ^\dag (x,t,\lambda^*) = \phi^{-1}  (x,t,-\lambda) ,
\end{split}\end{equation}
so for the scattering matrix we have
\begin{equation}\label{eq:Tdag'}\begin{split}
 T ^\dag (t,-\lambda^*) = T  (t,\lambda) ,
\end{split}\end{equation}
As a consequence for the Gauss factors we get:
\begin{equation}\label{eq:Spm'}\begin{aligned}
 T^-{}^\dag (-\lambda^*) & = \hat{S}^+(\lambda), &\qquad  T^+{}^\dag (-\lambda^*) & = \hat{S}^-(\lambda), \qquad
 D^\pm{}^\dag (\lambda^*) & = \hat{D}^\pm(-\lambda).
\end{aligned}\end{equation}
In analogy with the local reductions, the kernel of the resolvent has poles at the at the points $\lambda_2^\pm$ at which
$D^\pm (\lambda)$ have poles or zeroes. In particular, if $\lambda_2^+$ is an eigenvalue, then $-\lambda_2^+$ is also an
eigenvalue. For the reflection coefficients we obtain the constraints:
\begin{equation}\label{eq:tau'}\begin{split}
\tau^+(-\lambda) = -\rho^{+,*} (\lambda), \qquad  \tau^-(-\lambda) = -\rho^{-,*} (\lambda),
\end{split}\end{equation}

\section{Conclusion}

We demonstrated that the results concerning the GFT for nonlocal reductions
hold true also for the  MNLS cases, in particular for the Kulish-Sklyanin type models. The results are
natural extensions of the ones in \cite{GeSa} to the multicomponent cases.


\begin{thebibliography}{99}

\bibitem{AbBak}   M. Ablowitz, I. Bakirtas and  B. Ilan.
 "Wave collapse in a class of nonlocal nonlinear Schrddinger equation," Physica {\bf D 207}, 230—253, (2005).

\bibitem{AKNS} M.J. Ablowitz, D.J. Kaup, A.C.  Newell and H. Segur,
{\it The Inverse Scattering Transform-Fourier Analysis for Nonlinear
Problems},  Stud.  Appl. Math.  {\bf  53}, 249-315 (1974).

\bibitem{AblMusl} M. Ablowitz and Z. Musslimani, {\it Integrable Nonlocal Nonlinear Schr\"odinger
Equation}, Phys. Rev. Lett., {\bf 110} (2013) 064105(5).

\bibitem{AblMus2} Mark J. Ablowitz and Ziad H. Musslimani.
{\it Inverse scattering transform for the integrable nonlocal nonlinear Schr\"odinger equation}, under review (2015).

\bibitem{Konot} I.V. Barashenkov, D.A. Zezyulin, V.V. Konotop.
Exactly solvable Wadati potentials in the PT-symmetric Gross-Pitaevskii equation
{\bf arXiv preprint arXiv:1511.06633}.

\bibitem{Bender1} C. M. Bender and S. Boettcher, {\it Real Spectra in Non-hermitian Hamiltonians
Having ${\cal PT}$ Symmetry}, Phys. Rev. Lett 80 (1998) 5243--5246;\\
C. M. Bender, S. Boettcher and P. N. Meisinger, {\it ${\cal PT}$-Symmetric quantum Mechanics},  J. Math. Phys. {\bf 40} (1999) 2201--2229.

\bibitem{Bender2}  C. M. Bender, {\it Making Sense of Non-hermitian Hamiltonians},  Rep. Progr. Phys. {\bf 70} (2007) 947--1018 (E-print: {\tt hep-th/0703096}).

\bibitem{dwy08} Doktorov, E. V., Wang, J. and Yang, J.,
"Perturbation theory for bright spinor Bose-Einstein condensate solitons",
Phys. Rev. A {\bf 77}, 043617 (2008).

\bibitem{ForKu*83} A.~P.~Fordy and P.~P.~Kulish, {\it Nonlinear Schrodinger
Equations and Simple Lie Algebras}, Commun.\ Math.\ Phys.\ {\bf
89} (1983) 427--443.

\bibitem{G}   Gerdjikov V.~S.
\textit{ Generalized Fourier Transforms for the Soliton Equations.
Gauge Covariant Formulation.} Inverse Problems {\bf 2}, n.~1 (1986) 51--74.

\bibitem{TMF98} Gerdjikov V. S. \textit{Complete Integrability, Gauge
Equivalence and Lax Representations of the Inhomogeneous Nonlinear
Evolution Equations.} Theor. Math. Phys.   {\bf 92}  (1992)
374--386.

\bibitem{ConMath} V. S. Gerdjikov.
{\em Algebraic and Analytic Aspects of $N $-wave Type Equations.}
Contemporary Mathematics {\bf 301}, 35-68 (2002); {\bf nlin.SI/0206014}.

\bibitem{VSG2} V. S. Gerdjikov.
{\it Basic Aspects of Soliton Theory.} In: Eds.: I. M. Mladenov,
A. C. Hirshfeld. "Geometry, Integrability and Quantization", pp.
78-125; Softex, Sofia 2005. {\bf nlin.SI/0604004}

\bibitem{SIGMA-6} V. S. Gerdjikov, G. G. Grahovski. Multi-component NLS Models
on Symmetric Spaces: Spectral Properties versus Representations Theory.
SIGMA {\bf 6} (2010), 044, 29 pages; {\bf arXiv: 1006.0301 [nlin.SI]}.

\bibitem{TMF-144} V. S. Gerdjikov, G. G. Grahovski, N. A. Kostov.
{\it On the multi-component NLS type equations  on symmetric
spaces and their reductions.}
Theor. Math. Phys. {\bf 144} No. 2 1147-1156 (2005).

\bibitem{GKV} V.S. Gerdjikov, N.A. Kostov,  T.I. Valchev.
$N$-Wave Equations with Orthogonal Algebras: $\mathbb{Z}_2$ and
$\mathbb{Z}_2\times\mathbb{Z}_2$ Reductions and Soliton Solutions.
SIGMA {\bf 3}, paper 039 (2007); 19 pages. \qquad {\bf arXiv:nlin.SI/0703002}.

\bibitem{GKV1} V. S. Gerdjikov, N. A. Kostov, T. I. Valchev.
Solutions of multi-component NLS models and Spinor Bose-Einstein
condensates.  Physica D {\bf 238} 1306-1310 (2009)  
{\bf ArXiv:0802.4398 [nlin.SI]}.

\bibitem{SPIE-7501} V. S. Gerdjikov,  N. A. Kostov  and T. I. Valchev. {\it
Bose-Einstein condensates with $F=1$ and $F=2$.
Reductions and soliton interactions of multi-component NLS models.} In
(eds: Solomon M. Saltiel; Alexander A. Dreischuh; Ivan P. Christov)
Proceedings of SPIE   {\bf 7501}, 7501W (2009).  
{\bf arXiv: 1001.0168 [nlin.SI]}

\bibitem{1} V.~S.~Gerdjikov, G.~ G.~Grahovski, R.~I.~Ivanov and N.~A.~Kostov,
{\it $N $-wave interactions related to simple Lie algebras.
$\bbbz_2$- reductions and Soliton Solutions}, Inv. Problems {\bf 17} (2001) 999--1015.

\bibitem{gkv08} V. S. Gerdjikov, N. A. Kostov, T. I. Valchev.
{\it Solutions of multi-component NLS models and Spinor
Bose-Einstein condensates}, Physica D  {\bf 238} 1306-1310 (2009);  {\bf ArXiv:0802.4398 [nlin.SI]}.

\bibitem{GeSa}   V. S. Gerdjikov, A. Saxena.
Complete integrability of Nonlocal Nonlinear Schr\"odinger equation.
{\bf arXiv:1510.00480v1 [nlin.SI]}.

\bibitem{GGK05b} G.~G.~Grahovski, V.~S.~Gerdjikov, N.~A.~Kostov,
V.~A.~Atanasov, {\it New Integrable Multi-component NLS type
Equations on Symmetric Spaces: $Z_4$ and $Z_6$ reductions}, In
{\it ``Geometry, Integrability and Quantization VII''}, Eds: I.
Mladenov and M. De Leon, Softex, Sofia (2006), pp. 154--175.

\bibitem{Helg} S.~Helgasson,
{\it Differential Geometry, Lie Groups and Symmetric Spaces},
(Graduate studies in Mathematics, vol.34), AMS, Providence, Rhode
Island (2001).

\bibitem{I04} R.~I.~Ivanov, {\it On the dressing method for the
generalized Zakharov-Shabat system,}  Nucl. Phys. {\bf B 694},
(2004) 509--524.

\bibitem{Kevre*08} Nistazakis, H. E., Frantzeskakis, D. J., Kevrekidis, P. G., Malomed, B. A. and Carretero-Gonz´alez R.,
"Bright-Dark Soliton Complexes in Spinor Bose-Einstein Condensates" Phys. Rev. {\bf A 77},
033612 (2008).

\bibitem{SPIE-6604} N. A. Kostov, V. A. Atanasov, V. S. Gerdjikov, G. G. Grahovski.
{\it On the soliton solutions of the spinor Bose-Einstein condensate }.
Proceedings of SPIE {\bf 6604}, 66041T (2007).   Editors: Peter A. Atanasov, Tanja N. Dreischuh,
Sanka V. Gateva, Lubomir M. Kovachev.

\bibitem{KuSkl} P. P. Kulsh and E. K. Sklyanin.
$0(n)$-invariant nonlinear Schr\"odinger equation — a new completely integrable system.
Phys. Lett. A {\bf 84A} 349--352 (1981).

 \bibitem{Man}  Manakov, S.V.: On the theory of two-dimensional stationary
self-focusing of electromagnetic waves. Zh. Eksp. Teor. Fiz. \textbf{65},
 1392 (1973). (English translation) Sov. Phys. JETP \textbf{38}, 248 (1974). (in Russian)

\bibitem{Mikh} A.~V.~Mikhailov,
{\it The reduction problem and the inverse scattering problem},
Physica D {\bf 3} (1981) 73--117.

\bibitem{imww04}
J.~Ieda, T.~Miyakawa and M.~Wadati, {\it Matter-wave solitons in
an $F=1$ spinor Bose-Einstein condensate}, J. Phys. Soc. Jpn.
\textbf{73} (2004) 2996.

\bibitem{Ali1}  A. Mostafazadeh,    {\it Pseudo-hermiticity    versus     ${\cal PT}$-Symmetry I, II, III}, J. Math. Phys.
{\bf 43} (2002) 205--214 (E-print: {\tt math-ph/0107001}); 2814--2816 (E-print: {\tt math-ph/0110016}); 3944--3951 (E-print: {\tt math-ph/0203005}).

\bibitem{Ali2} A. Mostafazadeh, {\it Pseudo-hermiticity and Generalized ${\cal PT}$- and ${\cal CPT}$-Symmetries},
J. Math. Phys. {\bf 44} (2003) 974--989 (E-print: {\tt math-ph/0209018});\\
    A. Mostafazadeh, {\it Exact ${\cal PT}$-Symmetry Is Equivalent to Hermiticity}, J. Phys. A: Math. Gen.
    {\bf 36} (2003) 7081--7091 (E-print: {\tt quant-ph/0304080}).

\bibitem{PRA64} Nille N. Klausen, John L. Bohn and Chris H. Greene.
{\it Nature of spinor Bose-Einstein condensates in rubidium.}
Phys. Rev. A 64, 053602 (2001).

\bibitem{OM}
T.~Ohmi and K.~Machida, {\it Bose-Einstein condensation with
internal degrees of freedom in alkali atom gases} J. Phys. Soc.
Jpn. \textbf{67} (1998) 1822.

\bibitem{Sh} A. B. Shabat. \textit{The inverse scattering problem for a  system of differential equations.}
{}Functional Annal. \& Appl. {\bf 9}, n.3, 75 (1975) (In Russian);\\
A. B. Shabat. \textit{The inverse scattering problem.} Diff.  Equations {\bf 15}, 1824 (1979) (In Russian).

\bibitem{uiw06}
M.~Uchiyama, J.~Ieda and M.~Wadati, {\it Dark solitons in $F=1$
spinor Bose--Einstein condensate} J. Phys. Soc. Jpn. \textbf{75}
(2006) 064002.

\bibitem{uiw07} M.~Uchiyama, J.~Ieda, and M.~Wadati,
 {\it Multicomponent Bright Solitons in $F= 2$ Spinor Bose-Einstein Condensates}, J. Phys. Soc.
Japan,{\bf 76}, No. 7, (2007), 74005.

\bibitem{0710a} Uchino, Shun, Otsuka, Takaharu, Ueda, Masahito.
Dynamical symmetry in spinor Bose-Einstein condensates {\bf arXiv:0710.5210.}

\bibitem{UK}
M.~Ueda and M.~Koashi, {\it Theory of spin-2 Bose-Einstein
condensates: Spin correlations, magnetic response, and excitation
spectra}, Phys. Rev. A \textbf{65} (2002) 063602.

\bibitem{Val}  T. Valchev. {\it On a nonlocal nonlinear Schr\"odinger equation}, in  ``Mathematics in Industry",  Ed. A. Slavova,
Cambridge Scholars Publ., 2014, 36-52, ISBN 978-1-4438-6401-5 (Proc. of 8-th Annual Meeting of the
Bulgarian Section of SIAM, 18-19 December, 2013, Sofia, Bulgaria).

\bibitem{TV} T. I.  Valchev, {\it On Mikhailov's reduction group}, Phys. Lett A {\bf 379} (2015)  1877--1880.

\bibitem{ZMNP} V.~E.~Zakharov, S.~V.~Manakov, S.~P.~Novikov   and L.~I.~Pitaevskii,
{\it Theory of Solitons. The Inverse Scattering
Method}, Plenum Press (Consultant Bureau), N.Y., (1984).

\bibitem{ZaSh}  {V. E. Zakharov,   and A. B. Shabat}, {\it A scheme for integrating nonlinear evolution equations
of mathematical physics by the inverse scattering method. I \& II}, Funkts. Anal. Prilozhen.,
{\bf 8} (1974), 43--53; {\bf 13} (1979) no. 3, 13--22.

\bibitem{UFN}  A. A. Zyablovsky, A. P. Vinogradov, A. A. Pukhov, A. V. Dorofeenko and A A Lisyansky,
{\it ${\cal PT}$-symmetry in optics},  Phys.-Uspekhi {\bf 57} (2014), no. 11, 1063.


\end{thebibliography}
\end{document}